\documentclass[showpacs]{revtex4}
\begin{document}

\title{SUPPRESSING THE SPURIOUS STATES OF THE CENTRE OF MASS}
  
\author{P. Di\c t\u a and L. Micu}

\affiliation{Department of
Theoretical Physics\\ Horia Hulubei Institute for Physics and Nuclear Engineering\\ Bucharest POB MG-6, RO 077125}

\thanks{lmicu@theory.nipne.ro}


\begin{abstract}
Following Dirac's ideas concerning the quantization of constrained systems, we suggest to replace the free centre of mass Hamiltonian $H_{CM}$ by another operator which commutes with all the elements of the algebra generated via the commutation relations by $H_{CM}$ and the constraints which fix the centre of mass position. We show that the new Hamiltonian is a multiple of the identity operator and, as a result, its unique effect is to raise the internal energy levels by a constant amount.
\end{abstract}

\maketitle

\section*{}
In classical and nonrelativistic quantum mechanics (see e. g. \cite{am}) the first step in the standard approach to the two body bound state problem is to replace the individual variables $(\vec{r}_{1},~\vec{p}_{1}),~(\vec{r}_{2},~\vec{p}_{2})$ in the two-body Hamiltonian by the centre of mass variables $(\vec{R},\vec{P})$ and the relative ones $(\vec{r},\vec{p})$, where 
\begin{eqnarray}
&&\vec{R}=m_1/(m_1+m_2)~\vec{r}_1+m_2/(m_1+m_2)~\vec{r}_2\nonumber\\
&&\vec{P}=\vec{p}_1+\vec{p}_2\nonumber\\
&&\vec{r}=\vec{r}_1-\vec{r}_2\nonumber\\
&&\vec{p}=m_2/(m_1+m_2)~\vec{p}_1-m_1/(m_1+m_2)~\vec{p}_2\nonumber.
\end{eqnarray} 
Then, if the interaction potential $V$ depends only on the relative coordinates, the Hamiltonian separates in the free centre of mass part $H_{CM}$ and an internal Hamiltonian $H_{int}$:
\begin{equation}\label{H}
H=H_{CM}+H_{int}=\frac{\vec{P}^2}{2M}+\frac{\vec{p}^2}{2m_r}+V(r)
\end{equation}
where $M=m_1+m_2$ and $m_r=m_1m_2/(m_1+m_2)$.
By this mathematical trick the two-body problem separates in two independent single-particle problems for two fictious particles: the centre of mass problem and the relative problem. The first one is the problem of a free particle with the mass $M$ having the centre of mass coordinates and the second is the problem of a particle with the reduced mass in the potential well $V(r)$.  

We mention that the separation procedure \cite{am} applied to an $N$-body Hamiltonian gives a result similar to (\ref{H}). The centre of mass Hamiltonian has the same form as $H_{CM}$ where now $M=\sum_{i=1}^Nm_i$ and $H_{int}$ depends only on translational invariant (intrinsic) coordinates.
 
In the following we consider the centre of mass problem in the classical and quantum mechanics in the case of a bound system at rest, {\it i. e.} when the position of the centre of mass is fixed. 

In classical mechanics and Hamilton formulation, this condition takes the form 
\begin{equation}\label{cm}
\dot{\vec{R}}(t)=\{H_{CM},\vec{R}(t)\}=\frac{1}{M}\vec{P}=0
\end{equation} 
where $\{\cdot,\cdot\}$ is the Poisson bracket. This means that $\vec{R}(t)=\vec{R}_0$ is equivalent to $\vec{P}=0$ and hence the classical Hamiltonian which takes into account the  constraint reads
\begin{equation}\label{Hrest}
H=\frac{\vec{p}^2}{2m_r}+V(\vec{r}).
\end{equation}

In quantum mechanics the classical Hamiltonian $H$ is replaced by an operator ${\mathcal H}$ whose expression is obtained with the aid of the correspondence principle. If $\mathcal H$ does not explicitly depend on time the state of the physical system is described by a stationary wave function $\Psi(\vec{R},\vec{r})$. Following the separation procedure outlined before, it is the product of the centre of mass wave function $\psi_{CM}(\vec{R})$ and the internal wave function $\psi_{int}(\vec{r})$ which satisfy the Schr\"odinger equations 

\begin{equation}\label{cm1}
(H_{CM}-E_{CM})\psi_{CM}(\vec{R})=0
\end{equation}
and 
\begin{equation}
(H_{int}-E_{int})\psi_{int}(\vec{r})=0
\end{equation} 
with specific boundary conditions.  

Considering the centre of mass equation (\ref{cm1}) we notice the following: 

If, $\psi_{CM}(\vec{R})$ is the eigenfunction of $H_{CM}$, from the normalization condition it results that the probability to find the centre of mass in any finite volume $v$ is $v/(2\pi)^3V$ and tends to zero when $V\to\infty$, which does not agree with the real situation.  

On the contrary, if we require a certain localization and replace the condition $\vec{R}(t)=\vec{R}_0$ in classical mechanics by the confinement of the centre of mass to a 3-dimensional box of side $l$ centred in $\vec{R}_0$ the invariance at translations is destroyed. Moreover, solving the centre of mass equation (\ref{cm1}) with  Dirichlet boundary conditions we get an infinite set of states
\begin{equation}\label{fct}
\psi^{\{n\}}_{CM}(\vec{R})=\prod_{i=1}^3\psi_{n_1}(X-X_0)\psi_{n_2}(Y-Y_0)\psi_{n_3}(Z-Z_0)
\end{equation}
where $\psi_{n_i}(x)=\sin[n_i\pi(1/2+x/l)],~n_i=1,2,3....$ and the energy of the centre of mass is
\begin{equation}\label{ecm}
E_{CM}^{\{n\}}=\frac{\pi^2(n_1^2+n^2_2+n_3^2)}{2M~l^2}.
\end{equation}
We remark that the smaller is $l$, the higher is the kinetic energy of the centre of mass of a system at rest! 

Obviously, these states are the result of the incompatibility of the position and momentum observables in quantum mechanics, not of a dynamical scheme. They have no classical analog, have not been observed experimentaly, and hence are spurious states whose contribution must be eliminated from the final results.

In the two body case the centre of mass problem is usually ignored \cite{am}, the interest being focused on the internal wave function $\psi_{int}$ and on the internal energy levels $E_{int}$. 

The problem has been extensively studied in connection with the nuclear many body models \cite{rs}, where the achievement of a translational invariant, independent particle picture requires a clear separation of the centre of mass wave function and energy from the internal ones. The solution proposed by Lipkin et al. \cite{lst} is to introduce in the internal Hamiltonian the contribution of the redundant (or superfluous) coordinates which assure the independent treatment of the internal particles. These ones generate a new term in the internal Hamiltonian and new spurious solutions. The additional term in $H_{int}$ compensates the kinetic energy of the centre of mass in the approximation of equal masses but finally its contribution has to be subtracted from $E_{int}$. The separation of the spurious solutions requires special technics which, excepting the case of the harmonic oscillator, is rather hard (see, e.g., Ref.\, \cite{rs} (par. 11.3), \cite{lst}, \cite{l}, \cite{pt}). 

In this paper we propose a new solution. We are concerned only with the centre of mass Hamiltonian, so that the result is the same for any $N$-body bound systems. 

Taking into account Dirac's conclusions concerning the quantization of constrained systems, quoted in his books (see Ref. \cite{pamd}):

{\it (i)} The Hamiltonian of a constrained system is not uniquely defined;

{\it (ii)} The commutators (or the Poisson brackets) of the "naive" Hamiltonian with the constraints represent new constraints;

\noindent
we suggest to replace the free centre of mass Hamiltonian by another operator which is compatible with the constraints. 

As demonstrated by one of us in \cite{pd}, an operator of this kind exists if the constraints are compatible and if the Poisson-Lie algebra  generated by commutation or Poisson brackets from the naive Hamiltonian and the constraints is finite. It is a Casimir invariant of the algebra or a convenient element of its centre. This is the "right" Hamiltonian which has to be used in the quatization of constrained systems since it commutes with the constraints and hence assures their conservation in time. As it can be seen from some simple examples, the right Hamiltonian actually "absorbs" the constraints, so that these ones are identically satisfied by the solutions of the equation of motion.

It is important to notice that the procedures leading to the right Hamiltonian in classical and quantum mechanics are similar, so that the Hamiltonians obtained in this way are in perfect agreement. A clear example in this sense is the motion of a particle on a $n$ dimensional sphere \cite{pd} when the free Hamiltonian and the constraint read

\begin{equation}
H=\frac{1}{2m}\sum_{i=1}^{n+1}p_i^2,~~~~~U=\frac{1}{\vec{r}_0^2}\sum_{i=1}^{n+1}x_i^2-1=0.
\end{equation}
$H$, $U$ and their Poisson brackets (or their commutators) generate a closed algebra 
\begin{equation}\label{alg}
\{H,U\}=2V,~~\{H,V\}=\frac{2}{mr_0^2}H,~~\{V,U\}=\frac{2}{mr_0^2}U.
\end{equation}
which includes the new element 
\begin{equation}
V=\frac{1}{mr_0^2}\sum_{i=1}^{n+1}x_ip_i
\end{equation} 

In this particular case, the right Hamiltonian is the Casimir invariant of the Poisson-Lie algebra (\ref{alg}) and has the form
\begin{equation}
{\mathcal H}=(U+1)H-\frac{mr_0^2}{2}V^2=\frac{1}{2I}\sum_{i<j}^{n+1}L_{ij}^2
\end{equation}
where $L_{ij}=x_ip_j-x_jp_i$ are the components of the angular momentum and $I=mr_0^2$ is the momentum of inertia. One easily checks that $\{{\mathcal H},H\}=\{{\mathcal H},U\}=\{{\mathcal H},V\}=0$ which means that all the constraints are preserved in time. 

We apply the same procedure to the case of the centre of mass Hamiltonian. We consider the algebra generated by $H_{CM}$, the constraint $\vec{R}-\vec{R}_0=0$ and the elements obtained by commutation. We get the new elements
\begin{equation}\label{com}
[H_{CM},\vec{R}]=i\frac{1}{M}\hat{\vec{P}}
\end{equation}
and
\begin{equation}\label{M}
[\hat{\vec{P}},\vec{R}]=-i{\bf I}
\end{equation}
where ${\bf I}$ is the identity operator. The algebra closes under the commutation relation and has a centre made of the multiples of the identity operator which commute with all the elements. According to our prescription, a certain element having the dimension of energy of this centre is the right centre of mass Hamiltonian in the quantum mechanical framework. 

This is an expected result, because the algebra generated by $\hat{\vec{P}},~\vec{R}$ is irreducible and hence, according to Schur lemma, the only operator commuting with all its elements must be a multiple of ${\bf I}$. 

Then, in agreement with the classical result (\ref{Hrest}), the right Hamiltonian reads
\begin{equation}\label{Hnew}
{\mathcal H}=f(M)+H_{int}.
\end{equation}

From the absence of the centre of mass variables in (\ref{Hnew}) we infer that there is no centre of mass wave function and no spurious energy levels. The translational invariance is restored at the quantum level and the state of the bound system is described by the solution $\psi_{int}$ of the Schr\"odinger equation  
\begin{equation}\label{ecS}
\left(f(M)+H_{int}-{\mathcal E}\right)\psi_{int}(\vec{r})=0.
\end{equation}

In conclusion, the unique effect of the right Hamiltonian is to raise the energy levels of $H_{int}$ by a constant amount denoted by $f(M)$. This one cannot be directly measured, but from physical reasons one may suppose that $f(M)=M$. In this case ${\mathcal E}\to M$ when the strength of the interaction potential $V(\vec{r})$ tends to 0, and hence the negative eigenvalues $E_{int}$ of the internal Hamiltonian acquire the real meaning of a "mass defect" generated by the mutual attraction of the internal bodies.

The conclusion may be immediately extended to an isolated $N$-body system at rest where the right centre of mass Hamitonian is $H_{CM}=\sum_{i=1}^{N}m_i$. 

Closing, we mention that our result is the first step in the attempt to obtain a translational invariant, independent particle picture of bound systems where the place of the redundant variables is taken by the effective variables of the internal field. This makes the subject of a forthcoming paper. 

\begin{acknowledgments} The authors thank Fl. Stancu and H. Scutaru for valuable remarks and encouragement. The financial support from the Romanian Ministry of Education and Research under the contract PN06 35 01 01 is acknowledged.
\end{acknowledgments}

\end{document}